\documentclass{article}
\pdfoutput=1
\usepackage[utf8]{inputenc}
\usepackage[tiny]{titlesec}
\usepackage{latexsym,nicefrac,multirow}

\usepackage{savetrees}
\usepackage{filecontents}

\usepackage[colorlinks=true,allcolors=blue]{hyperref}

\usepackage[backend=bibtex,natbib=true,style=numeric,maxbibnames=99,firstinits=true]{biblatex}
\renewbibmacro{in:}{%
  \ifentrytype{article}{}{%
  \printtext{\bibstring{in}\intitlepunct}}}

\date{} 

\usepackage{fancyvrb,relsize}
\fvset{frame=single,framerule=.2mm,fontsize=\relscale{.6}}
\input{lst/pygments}

\newcommand*\FancyVerbStartString{}
\newcommand*\FancyVerbStopString{}

\usepackage{float}
\newfloat{Listing}{hbtp}{lop}

\newcommand{\prog}[1]{{\tt#1}}

\usepackage{authblk}
\author[1,2]{Dorota Jarecka}
\author[1]{Sylwester Arabas}
\author[2]{Davide Del Vento}
\affil[1]{Institute of Geophysics, Faculty of Physics, University of Warsaw, Poland}
\affil[2]{National Center for Atmospheric Research, USA}

\title{Python bindings for libcloudph++}

\begin{document}
\twocolumn[
  \begin{@twocolumnfalse}
    \maketitle
    \begin{abstract}
      This technical note introduces the Python bindings
        for {\em libcloudph++}.
      The~{\em libcloudph++} is a C++ library of algorithms
        for~representing atmospheric cloud microphysics in numerical models.
      The~bindings expose the complete functionality of~the~library
        to the Python users.
      The~bindings are implemented using the {\em Boost.Python} C++
        library and use {\em NumPy} arrays.
      This~note includes listings with Python scripts exemplifying the use
        of selected library components.
      An~example solution for using the~Python bindings to access {\em libcloudph++}
        from Fortran is presented.
    \end{abstract}
    \vspace{2em}
  \end{@twocolumnfalse}
]

  \section{Introduction}

  This paper describes how to use the {\em libcloudph++} from 
    the~Python\footnote{\url{http://python.org/}} programming language.
  The {\em libcloudph++}\footnote{\url{http://libcloudphxx.igf.fuw.edu.pl/}} 
    is a free and open-source C++ library 
    of algorithms for representing cloud microphysics in 
    atmospheric numerical models.
  A~detailed description of the library and its C++ interface 
    has been described in \cite{Arabas_et_al_2015}.
  In~short, the library covers three numerical schemes describing 
    processes occurring in warm clouds (i.e. in the absence of~ice).
  The~represented processes cover cloud-droplet condensational growth
    and formation of rain drops through collisions and coalescence.
  The~first implemented scheme is a simplistic, so-called single-moment
    bulk scheme that allows predicting the~total mass of cloud water
    and of rain water in a volume of air.
  The~second scheme is a double-moment bulk scheme that adds 
    prediction of~the~number concentration of cloud droplets and rain drops.
  The~third scheme is based on the concept of particle tracking.
  The~particle-based scheme implemented in {\em libcloudph++} represents
    collisional growth of particles using a probabilistic Monte-Carlo type
    model.
  Furthermore, it is implemented for use on both multiple CPU threads as well
    as on a GPU.

  Access to {\em libcloudph++} from Python is provided through so-called bindings.
  The bindings to {\em libcloudph++} allow using the~library 
    from Python, without requiring the user to interact with the native 
    C++ interface.
  The Python bindings significantly facilitate the use of the library and
    add relatively little runtime overhead (particularly in the case 
    of the resource-intensive particle-based scheme).
  Python has simpler syntax than C++, its~philosophy emphasises succinct code 
    (see \cite{Arabas_et_al_2014} for a geoscience-relevant case study comparing Python, C++ and Fortran).
  Moreover, Python is widespread across the atmospheric science community \cite{Lin_2012b}. 
  The~vast availability of software packages for interfacing
    Python codes from other languages \cite{Langtangen_2008}
    makes the Python bindings a good starting point for 
    using {\em libcloudph++} from other languages, for instance from Fortran.
  Arguably, the embraced approach makes the best out of salient features of two languages by using:
  \begin{description}
    \item[C++]{for implementing numerically-intensive concurrency-enabled algorithms for both CPU and GPU, and encapsulating them in a library;}
    \item[Python]{for equipping the library with rapid-development features and for interfacing with other languages.}
  \end{description}
  This note is intended as a companion to the documentation of the library
    presented in \cite{Arabas_et_al_2015} and is structured as follows.
  Section~\ref{sec:python} presents the programming interface of the Python bindings.
  It~includes examples of Python code with calls to two out of four components
    of the library, namely the commons (section~\ref{sec:commons}) 
    and the single-moment bulk scheme (section~\ref{sec:blk_1m}).
  Section~\ref{sec:fortran} exemplifies how to use the Python bindings 
    to call the {\em libcloudph++} from Fortran.
  Appendix~\ref{sec:install} describes how to obtain and install {\em libcloudph++}
    and the Python bindings.
  
  \section{Summary of the Python interface}\label{sec:python}
 
  \begin{table*}
    \caption{\label{tab:summary}
      Contents of the \prog{libcloudphxx} Python package. 
      The \prog{a.$<$b,c,d$>$} notation indicates that \prog{b},\prog{c},\prog{d} are all attributes of~\prog{a}.
      The~\prog{blk\_2m}~and~\prog{lgrngn} modules listed at the bottom of the table are part of the bindings but
        are not described in this note.
    }
    \footnotesize
    \center
    \begin{tabular}{lccc}
       item
         & type   
           & summary         
             & example                     
               \\ \hline\hline
       \prog{git\_revision}  
         & string 
           & version number  
             & Lst.~\ref{lst:git_revision} 
               \\ \hline
       \prog{common}
         & module 
           &
             &  
               \\
       \prog{common.$<$R\_v, R\_d, eps, c\_pd, c\_pv, g, p\_1000, rho\_w$>$} 
         & floats 
           & physical constants 
             & Lst.~\ref{lst:constants} 
               \\
       \prog{common.th\_std2dry(th, rv)} 
         & function
           & converts $\theta$ to $\theta_d$
             & Lst.~\ref{lst:th_drystd}
               \\
       \prog{common.th\_dry2std(th, rv)} 
         & function
           & converts $\theta_d$ to $\theta$
             & Lst.~\ref{lst:th_drystd}
               \\
       \prog{common.T(th\_d, rh\_d)} 
         & function 
           & implements Eq.~A14 in \cite{Arabas_et_al_2015} 
             & Lst.~\ref{lst:th_drystd}
               \\
       \prog{common.p(rh, rv, T)} 
         & function 
           & implements Eq.~A15 in \cite{Arabas_et_al_2015}
             & Lst.~\ref{lst:th_drystd}
               \\
       \prog{common.p\_vs(T)} & function & implements Eq.~15 in \cite{Arabas_and_Pawlowska_2011} & Lst.~\ref{lst:p_vs} \\
       \prog{common.rw3\_cr(rd3, kappa, T)}  & function & critical radius cubed \cite{Petters_et_al_2007} & Lst.~\ref{lst:kappa_crit},\ref{lst:test.py},\ref{lst:test.f}\\
       \prog{common.S\_cr(rd3, kappa, T)}    & function & critical saturation \cite{Petters_et_al_2007} & Lst.~\ref{lst:kappa_crit},\ref{lst:test.py},\ref{lst:test.f} \\ \hline
       \prog{blk\_1m} & module &  &  \\
       \prog{blk\_1m.opts\_t} & class & scheme options & Lst.~\ref{lst:blk_1m_opts} \\
       \prog{blk\_1m.opts\_t.$<$cond,cevp,revp,conv,accr,sedi$>$} & bools & process toggling flags & Lst.~\ref{lst:blk_1m_opts} \\
       \prog{blk\_1m.opts\_t.r\_c0} & float & autoconversion threshold & Lst.~\ref{lst:blk_1m_opts} \\
       \prog{blk\_1m.opts\_t.r\_eps} & float & saturation adjustment tolerance & Lst.~\ref{lst:blk_1m_opts} \\
       \prog{blk\_1m.adj\_cellwise(opts, $\rho_d$, $\theta_d$, $r_v$, $r_c$, $r_r$, $\Delta t$)} & function & condensation & Lst.~\ref{lst:blk_1m_adj} \\
       \prog{blk\_1m.rhs\_cellwise(opts, $\dot{r}_c$, $\dot{r}_r$, $r_c$, $r_r$)} & function & coalescence &  \\
       \prog{blk\_1m.rhs\_columnwise(opts, $\dot{r}_r$, $\rho_d$, $r_r$, $\Delta z$)} & function & sedimentation &  \\ \hline
       \prog{blk\_2m} & module & & \\ 
       \ldots         &        & & \\ \hline
       \prog{lgrngn} & module & & \\
       \ldots         &        & & \\ \hline
    \end{tabular}
  \end{table*}

  All elements of the Python bindings for {\em libcloudph++} are contained in 
    the \prog{libcloudphxx} Python package.
  Naming of the package components closely follows the native C++ interface.
  The~Python package contents are summarised in
    Table~\ref{tab:summary} and~described through examples in the
    following subsections.

  The current version of the package is compatible with Python~2 only.
  The Python interface uses {\em NumPy}\footnote{\url{http://numpy.org/}} arrays.
  Error handling is carried out by translating C++ exceptions into Python
    \mbox{\prog{RuntimeError}} exceptions.
  The bindings are implemented in~C++ using the~{\em Boost.Python}\footnote{\url{http://boost.org/libs/python/}}
    C++ library \cite{Abrahams_and_Grosse-Kunstleve_2003}.

  After successful installation (see appendix~\ref{sec:install}), the program given in
    Listing~\ref{lst:git_revision} will print the version number of the library
    expressed as a git revision id.
  \begin{Listing}
  \refstepcounter{Listing}%
  \fvset{label=Listing~\theListing}%
  \fvset{gobble=0}%
  \renewcommand*\FancyVerbStartString{\PY{c}{\PYZsh{}\PYZlt{}listing\PYZhy{}1\PYZgt{}}}%
  \renewcommand*\FancyVerbStopString{\PY{c}{\PYZsh{}\PYZlt{}/listing\PYZhy{}1\PYZgt{}}}%
  \begin{Verbatim}[commandchars=\\\{\}]
\PY{c}{\PYZsh{}\PYZlt{}listing\PYZhy{}1\PYZgt{}}
\PY{k+kn}{import} \PY{n+nn}{libcloudphxx}
\PY{k}{print} \PY{n}{libcloudphxx}\PY{o}{.}\PY{n}{git\PYZus{}revision}
\PY{c}{\PYZsh{}\PYZlt{}/listing\PYZhy{}1\PYZgt{}}
\end{Verbatim}
  \vspace{-1.5em}%

    \label{lst:git_revision}
  \end{Listing}
  
  \subsection{Commons}\label{sec:commons}

  The \prog{libcloudphxx.common} module contains a collection of physical constants
    and formul\ae.
  The constants exposed through the Python bindings can be printed
    with the instructions in Listing~\ref{lst:constants}.
  \begin{Listing}
  \refstepcounter{Listing}%
  \fvset{label=Listing~\theListing}%
  \fvset{gobble=0}%
  \renewcommand*\FancyVerbStartString{\PY{c}{\PYZsh{}\PYZlt{}listing\PYZhy{}1\PYZgt{}}}%
  \renewcommand*\FancyVerbStopString{\PY{c}{\PYZsh{}\PYZlt{}/listing\PYZhy{}1\PYZgt{}}}%
  \begin{Verbatim}[commandchars=\\\{\}]
\PY{c}{\PYZsh{}\PYZlt{}listing\PYZhy{}1\PYZgt{}}
\PY{k+kn}{from} \PY{n+nn}{libcloudphxx} \PY{k+kn}{import} \PY{n}{common}
\PY{k}{print} \PY{l+s}{\PYZdq{}}\PY{l+s}{R\PYZus{}d:}\PY{l+s}{\PYZdq{}}\PY{p}{,}    \PY{n}{common}\PY{o}{.}\PY{n}{R\PYZus{}d}    \PY{c}{\PYZsh{} gas constant for dry air}
\PY{k}{print} \PY{l+s}{\PYZdq{}}\PY{l+s}{R\PYZus{}v:}\PY{l+s}{\PYZdq{}}\PY{p}{,}    \PY{n}{common}\PY{o}{.}\PY{n}{R\PYZus{}v}    \PY{c}{\PYZsh{} gas constant for water vapour}
\PY{k}{print} \PY{l+s}{\PYZdq{}}\PY{l+s}{eps:}\PY{l+s}{\PYZdq{}}\PY{p}{,}    \PY{n}{common}\PY{o}{.}\PY{n}{eps}    \PY{c}{\PYZsh{} ratio of the above}
\PY{k}{print} \PY{l+s}{\PYZdq{}}\PY{l+s}{c\PYZus{}pd:}\PY{l+s}{\PYZdq{}}\PY{p}{,}   \PY{n}{common}\PY{o}{.}\PY{n}{c\PYZus{}pd}   \PY{c}{\PYZsh{} specific heat of dry air}
\PY{k}{print} \PY{l+s}{\PYZdq{}}\PY{l+s}{c\PYZus{}pv:}\PY{l+s}{\PYZdq{}}\PY{p}{,}   \PY{n}{common}\PY{o}{.}\PY{n}{c\PYZus{}pv}   \PY{c}{\PYZsh{} specific heat of water vapour}
\PY{k}{print} \PY{l+s}{\PYZdq{}}\PY{l+s}{g:}\PY{l+s}{\PYZdq{}}\PY{p}{,}      \PY{n}{common}\PY{o}{.}\PY{n}{g}      \PY{c}{\PYZsh{} acceleration due to gravity}
\PY{k}{print} \PY{l+s}{\PYZdq{}}\PY{l+s}{p\PYZus{}1000:}\PY{l+s}{\PYZdq{}}\PY{p}{,} \PY{n}{common}\PY{o}{.}\PY{n}{p\PYZus{}1000} \PY{c}{\PYZsh{} reference pressure of 1000 hPa}
\PY{k}{print} \PY{l+s}{\PYZdq{}}\PY{l+s}{rho\PYZus{}w:}\PY{l+s}{\PYZdq{}}\PY{p}{,}  \PY{n}{common}\PY{o}{.}\PY{n}{rho\PYZus{}w}  \PY{c}{\PYZsh{} density of water}
\PY{c}{\PYZsh{}\PYZlt{}/listing\PYZhy{}1\PYZgt{}}
\end{Verbatim}
  \vspace{-1.5em}%

    \label{lst:constants}
  \end{Listing}

  \begin{Listing}
  \refstepcounter{Listing}%
  \fvset{label=Listing~\theListing}%
  \fvset{gobble=0}%
  \renewcommand*\FancyVerbStartString{\PY{c}{\PYZsh{}\PYZlt{}listing\PYZhy{}1\PYZgt{}}}%
  \renewcommand*\FancyVerbStopString{\PY{c}{\PYZsh{}\PYZlt{}/listing\PYZhy{}1\PYZgt{}}}%
  \begin{Verbatim}[commandchars=\\\{\}]
\PY{c}{\PYZsh{}\PYZlt{}listing\PYZhy{}1\PYZgt{}}
\PY{k+kn}{from} \PY{n+nn}{libcloudphxx} \PY{k+kn}{import} \PY{n}{common}

\PY{n}{tht} \PY{o}{=} \PY{l+m+mi}{300} \PY{c}{\PYZsh{} K}
\PY{n}{r\PYZus{}v} \PY{o}{=} \PY{o}{.}\PY{l+m+mo}{01} \PY{c}{\PYZsh{} kg/kg}

\PY{n}{tht\PYZus{}d} \PY{o}{=} \PY{n}{common}\PY{o}{.}\PY{n}{th\PYZus{}std2dry}\PY{p}{(}\PY{n}{tht}\PY{p}{,} \PY{n}{r\PYZus{}v}\PY{p}{)}
\PY{k}{assert}\PY{p}{(}\PY{n}{tht} \PY{o}{==} \PY{n}{common}\PY{o}{.}\PY{n}{th\PYZus{}dry2std}\PY{p}{(}\PY{n}{tht\PYZus{}d}\PY{p}{,} \PY{n}{r\PYZus{}v}\PY{p}{)}\PY{p}{)}

\PY{n}{rho\PYZus{}d} \PY{o}{=} \PY{l+m+mi}{1} \PY{c}{\PYZsh{} kg / m3}
\PY{n}{T} \PY{o}{=} \PY{n}{common}\PY{o}{.}\PY{n}{T}\PY{p}{(}\PY{n}{tht\PYZus{}d}\PY{p}{,} \PY{n}{rho\PYZus{}d}\PY{p}{)}
\PY{n}{p} \PY{o}{=} \PY{n}{common}\PY{o}{.}\PY{n}{p}\PY{p}{(}\PY{n}{rho\PYZus{}d}\PY{p}{,} \PY{n}{r\PYZus{}v}\PY{p}{,} \PY{n}{T}\PY{p}{)}
\PY{c}{\PYZsh{}\PYZlt{}/listing\PYZhy{}1\PYZgt{}}
\PY{k}{print} \PY{n}{T}\PY{p}{,}\PY{n}{p}
\end{Verbatim}
  \vspace{-1.5em}%

    \label{lst:th_drystd}
  \end{Listing}
  As of the current release, there are seven formul\ae~available in~the~\prog{common}
    module of the Python bindings, see Table.~\ref{tab:summary}.
  The~first four functions
    convert the thermodynamic variables used in {\em libcloudph++}
    (dry-air potential temperature $\theta_d$, dry-air density $\rho_d$, water vapour mixing ration $r_v$) 
    to other commonly used variables
    (for details, see appendix~A in \cite{Arabas_et_al_2015}).
  Listing~\ref{lst:th_drystd} presents example use of two functions 
    for converting between the~$\theta_d$ and~the standard potential temperature.
  It also depicts how to use the functions for diagnosing the temperature~$T$ as~a~function 
    of~$\theta_d$ and~$\rho_d$, and for diagnosing the pressure $p$ as a 
    function of $\rho_d$, $r_v$ and $T$ (eqs.~A14 and A15 in \cite{Arabas_et_al_2015}).

  The \prog{p\_vs(T)} function calculates the saturated vapour pressure as a function of temperature
    using an analytic solution to the Clausius-Clapeyron equation
    used in the library.
  Listing~\ref{lst:p_vs} depicts example use of this function to calculate 
    the boiling temperature of water for atmospheric pressure of 500~hPa.
  Output of the program is given in Listing~\ref{lst:p_vs_out}.
  \begin{Listing}
  \refstepcounter{Listing}%
  \fvset{label=Listing~\theListing}%
  \fvset{gobble=0}%
  \renewcommand*\FancyVerbStartString{\PY{c}{\PYZsh{}\PYZlt{}listing\PYZhy{}1\PYZgt{}}}%
  \renewcommand*\FancyVerbStopString{\PY{c}{\PYZsh{}\PYZlt{}/listing\PYZhy{}1\PYZgt{}}}%
  \begin{Verbatim}[commandchars=\\\{\}]
\PY{c}{\PYZsh{}\PYZlt{}listing\PYZhy{}1\PYZgt{}}
\PY{k+kn}{from} \PY{n+nn}{libcloudphxx} \PY{k+kn}{import} \PY{n}{common}
\PY{k+kn}{from} \PY{n+nn}{scipy} \PY{k+kn}{import} \PY{n}{optimize}

\PY{n}{p\PYZus{}atm} \PY{o}{=} \PY{l+m+mi}{50000} \PY{c}{\PYZsh{} Pa}

\PY{c}{\PYZsh{} p\PYZus{}atm = p\PYZus{}vs(T\PYZus{}boil)}
\PY{k}{def} \PY{n+nf}{fun}\PY{p}{(}\PY{n}{T}\PY{p}{)}\PY{p}{:}
  \PY{k}{return} \PY{n}{p\PYZus{}atm} \PY{o}{\PYZhy{}} \PY{n}{common}\PY{o}{.}\PY{n}{p\PYZus{}vs}\PY{p}{(}\PY{n}{T}\PY{p}{[}\PY{l+m+mi}{0}\PY{p}{]}\PY{p}{)}

\PY{n}{zero} \PY{o}{=} \PY{l+m+mf}{273.15}
\PY{n}{res} \PY{o}{=} \PY{n}{optimize}\PY{o}{.}\PY{n}{root}\PY{p}{(}\PY{n}{fun}\PY{p}{,} \PY{n}{zero}\PY{o}{+}\PY{l+m+mi}{100}\PY{p}{)} \PY{c}{\PYZsh{} first guess: 100 C}
\PY{k}{assert}\PY{p}{(}\PY{n}{res}\PY{o}{.}\PY{n}{success}\PY{p}{)}
\PY{k}{print} \PY{l+s}{\PYZdq{}}\PY{l+s}{T\PYZus{}boil @ \PYZob{}0:g\PYZcb{} hPa: \PYZob{}1:g\PYZcb{} C}\PY{l+s}{\PYZdq{}}\PY{o}{.}\PY{n}{format}\PY{p}{(}
  \PY{n}{p\PYZus{}atm} \PY{o}{/} \PY{l+m+mi}{100}\PY{p}{,} 
  \PY{n}{res}\PY{o}{.}\PY{n}{x}\PY{p}{[}\PY{l+m+mi}{0}\PY{p}{]} \PY{o}{\PYZhy{}} \PY{n}{zero}
\PY{p}{)}
\PY{c}{\PYZsh{}\PYZlt{}/listing\PYZhy{}1\PYZgt{}}
\end{Verbatim}
  \vspace{-1.5em}%

    \label{lst:p_vs}
  \end{Listing}
  \begin{Listing}
  \refstepcounter{Listing}%
  \fvset{label=Listing~\theListing}%
  \fvset{gobble=0}%
  \renewcommand*\FancyVerbStartString{\PYZsh{}\PYZlt{}listing\PYZhy{}1\PYZgt{}}%
  \renewcommand*\FancyVerbStopString{\PYZsh{}\PYZlt{}/listing\PYZhy{}1\PYZgt{}}%
  \begin{Verbatim}[commandchars=\\\{\}]
\PYZsh{}\PYZlt{}listing\PYZhy{}1\PYZgt{}
T\PYZus{}boil @ 500 hPa: 81.7841 C
\PYZsh{}\PYZlt{}/listing\PYZhy{}1\PYZgt{}
\end{Verbatim}
  \vspace{-1.5em}%

    \label{lst:p_vs_out}
  \end{Listing}

  Note that the formul\ae~implemented in the \prog{common} module accept and
    return only double-precision scalars, hence the need to use \prog{T[0]}
    in the definition of \prog{fun(T)} in Listing~\ref{lst:p_vs} 
    (the SciPy\footnote{\url{http://scipy.org/}} 
    \prog{root} routine uses {\em NumPy} arrays even for single-equation problems).

  The last two functions available in the \prog{common} module
    compute the critical radius and the critical 
    saturation (see chapter~5 in~\cite{Curry_and_Webster_1999}) 
    using the kappa-K\"ohler parameterisation of hygroscopicity of
    water-solution droplets \cite{Petters_et_al_2007}.
  Listing~\ref{lst:kappa_crit} shows an example Python script generating
    a table of values of critical radius and~supersaturation for five
    different nucleus radii (compare Table~5.1 in~\cite{Curry_and_Webster_1999}).
  Output of the script is given in Listing~\ref{lst:kappa_crit_out}.
  \begin{Listing}
  \refstepcounter{Listing}%
  \fvset{label=Listing~\theListing}%
  \fvset{gobble=0}%
  \renewcommand*\FancyVerbStartString{\PY{c}{\PYZsh{}\PYZlt{}listing\PYZhy{}1\PYZgt{}}}%
  \renewcommand*\FancyVerbStopString{\PY{c}{\PYZsh{}\PYZlt{}/listing\PYZhy{}1\PYZgt{}}}%
  \begin{Verbatim}[commandchars=\\\{\}]
\PY{c}{\PYZsh{}\PYZlt{}listing\PYZhy{}1\PYZgt{}}
\PY{k+kn}{from} \PY{n+nn}{libcloudphxx.common} \PY{k+kn}{import} \PY{n}{rw3\PYZus{}cr}\PY{p}{,} \PY{n}{S\PYZus{}cr}

\PY{n}{kappa} \PY{o}{=} \PY{l+m+mf}{1.28} \PY{c}{\PYZsh{} [1]}
\PY{n}{T}     \PY{o}{=} \PY{l+m+mi}{273}  \PY{c}{\PYZsh{} [K]}

\PY{k}{print} \PY{l+s}{\PYZsq{}}\PY{l+s}{\PYZob{}0: \PYZgt{}10\PYZcb{}\PYZob{}1: \PYZgt{}10\PYZcb{}\PYZob{}2: \PYZgt{}10\PYZcb{}}\PY{l+s}{\PYZsq{}}\PY{o}{.}\PY{n}{format}\PY{p}{(}
  \PY{l+s}{\PYZsq{}}\PY{l+s}{rd [um]}\PY{l+s}{\PYZsq{}}\PY{p}{,}\PY{l+s}{\PYZsq{}}\PY{l+s}{r* [um]}\PY{l+s}{\PYZsq{}}\PY{p}{,}\PY{l+s}{\PYZsq{}}\PY{l+s}{S*\PYZhy{}1 [}\PY{l+s}{\PYZpc{}}\PY{l+s}{]}\PY{l+s}{\PYZsq{}}
\PY{p}{)}
\PY{k}{for} \PY{n}{rd} \PY{o+ow}{in} \PY{p}{(}\PY{o}{.}\PY{l+m+mo}{0223}\PY{p}{,} \PY{o}{.}\PY{l+m+mo}{047}\PY{l+m+mi}{9}\PY{p}{,} \PY{o}{.}\PY{l+m+mi}{103}\PY{p}{,} \PY{o}{.}\PY{l+m+mi}{223}\PY{p}{,} \PY{o}{.}\PY{l+m+mi}{479}\PY{p}{)}\PY{p}{:} \PY{c}{\PYZsh{} um}
  \PY{n}{rd3} \PY{o}{=} \PY{n+nb}{pow}\PY{p}{(}\PY{n}{rd} \PY{o}{*} \PY{l+m+mf}{1e\PYZhy{}6}\PY{p}{,} \PY{l+m+mi}{3}\PY{p}{)}
  \PY{k}{print} \PY{l+s}{\PYZsq{}}\PY{l+s}{\PYZob{}0: \PYZgt{}10g\PYZcb{}\PYZob{}1: \PYZgt{}10.2g\PYZcb{}\PYZob{}2: \PYZgt{}10.2g\PYZcb{}}\PY{l+s}{\PYZsq{}}\PY{o}{.}\PY{n}{format}\PY{p}{(}
    \PY{n}{rd}\PY{p}{,} 
    \PY{p}{(}\PY{n}{rw3\PYZus{}cr}\PY{p}{(}\PY{n}{rd3}\PY{p}{,} \PY{n}{kappa}\PY{p}{,} \PY{n}{T}\PY{p}{)}\PY{o}{*}\PY{o}{*}\PY{p}{(}\PY{l+m+mi}{1}\PY{o}{/}\PY{l+m+mf}{3.}\PY{p}{)}\PY{p}{)}\PY{o}{*}\PY{l+m+mf}{1e6}\PY{p}{,}
    \PY{p}{(}\PY{n}{S\PYZus{}cr}\PY{p}{(}\PY{n}{rd3}\PY{p}{,} \PY{n}{kappa}\PY{p}{,} \PY{n}{T}\PY{p}{)} \PY{o}{\PYZhy{}} \PY{l+m+mi}{1}\PY{p}{)} \PY{o}{*} \PY{l+m+mi}{100}
  \PY{p}{)}
\PY{c}{\PYZsh{}\PYZlt{}/listing\PYZhy{}1\PYZgt{}}
\end{Verbatim}
  \vspace{-1.5em}%

    \label{lst:kappa_crit}
  \end{Listing}
  \begin{Listing}
  \refstepcounter{Listing}%
  \fvset{label=Listing~\theListing}%
  \fvset{gobble=0}%
  \renewcommand*\FancyVerbStartString{\PYZsh{}\PYZlt{}listing\PYZhy{}1\PYZgt{}}%
  \renewcommand*\FancyVerbStopString{\PYZsh{}\PYZlt{}/listing\PYZhy{}1\PYZgt{}}%
  \begin{Verbatim}[commandchars=\\\{\}]
\PYZsh{}\PYZlt{}listing\PYZhy{}1\PYZgt{}
   rd [um]   r* [um]  S*\PYZhy{}1 [\PYZpc{}]
    0.0223      0.19      0.39
    0.0479      0.61      0.13
     0.103       1.9      0.04
     0.223       6.1     0.012
     0.479        19     0.004
\PYZsh{}\PYZlt{}/listing\PYZhy{}1\PYZgt{}
\end{Verbatim}
  \vspace{-1.5em}%

    \label{lst:kappa_crit_out}
  \end{Listing}

  \subsection{Single-moment bulk scheme}\label{sec:blk_1m}

  Access to the single-moment bulk scheme implemented in {\em libcloudph++} 
    is provided through the \prog{libcloudphxx.blk\_1m} module.
  The single-moment scheme 
    extends the set of model state variables by adding two mass mixing ratios,
    namely the cloud water mixing ratio $r_c$ and the rain water mixing radio $r_r$.

  Options of the scheme that can be altered at runtime are grouped as attributes of the \prog{blk\_1m.opts\_t} class.
  The default values of all options are set upon creating an instance of 
    \prog{opts\_t}, see Listing~\ref{lst:blk_1m_opts}.
  \begin{Listing}
  \refstepcounter{Listing}%
  \fvset{label=Listing~\theListing}%
  \fvset{gobble=0}%
  \renewcommand*\FancyVerbStartString{\PY{c}{\PYZsh{}\PYZlt{}listing\PYZhy{}1\PYZgt{}}}%
  \renewcommand*\FancyVerbStopString{\PY{c}{\PYZsh{}\PYZlt{}/listing\PYZhy{}1\PYZgt{}}}%
  \begin{Verbatim}[commandchars=\\\{\}]
\PY{c}{\PYZsh{}\PYZlt{}listing\PYZhy{}1\PYZgt{}}
\PY{k+kn}{from} \PY{n+nn}{libcloudphxx} \PY{k+kn}{import} \PY{n}{blk\PYZus{}1m}

\PY{n}{opts} \PY{o}{=} \PY{n}{blk\PYZus{}1m}\PY{o}{.}\PY{n}{opts\PYZus{}t}\PY{p}{(}\PY{p}{)}
\PY{k}{print} \PY{l+s}{\PYZdq{}}\PY{l+s}{cond  =}\PY{l+s}{\PYZdq{}}\PY{p}{,} \PY{n}{opts}\PY{o}{.}\PY{n}{cond}
\PY{k}{print} \PY{l+s}{\PYZdq{}}\PY{l+s}{cevp  =}\PY{l+s}{\PYZdq{}}\PY{p}{,} \PY{n}{opts}\PY{o}{.}\PY{n}{cevp}
\PY{k}{print} \PY{l+s}{\PYZdq{}}\PY{l+s}{revp  =}\PY{l+s}{\PYZdq{}}\PY{p}{,} \PY{n}{opts}\PY{o}{.}\PY{n}{revp} 
\PY{k}{print} \PY{l+s}{\PYZdq{}}\PY{l+s}{conv  =}\PY{l+s}{\PYZdq{}}\PY{p}{,} \PY{n}{opts}\PY{o}{.}\PY{n}{conv} 
\PY{k}{print} \PY{l+s}{\PYZdq{}}\PY{l+s}{accr  =}\PY{l+s}{\PYZdq{}}\PY{p}{,} \PY{n}{opts}\PY{o}{.}\PY{n}{accr} 
\PY{k}{print} \PY{l+s}{\PYZdq{}}\PY{l+s}{sedi  =}\PY{l+s}{\PYZdq{}}\PY{p}{,} \PY{n}{opts}\PY{o}{.}\PY{n}{sedi} 
\PY{k}{print} \PY{l+s}{\PYZdq{}}\PY{l+s}{r\PYZus{}c0  =}\PY{l+s}{\PYZdq{}}\PY{p}{,} \PY{n}{opts}\PY{o}{.}\PY{n}{r\PYZus{}c0}
\PY{k}{print} \PY{l+s}{\PYZdq{}}\PY{l+s}{r\PYZus{}eps =}\PY{l+s}{\PYZdq{}}\PY{p}{,} \PY{n}{opts}\PY{o}{.}\PY{n}{r\PYZus{}eps}

\PY{c}{\PYZsh{} ...}
\PY{c}{\PYZsh{}\PYZlt{}/listing\PYZhy{}1\PYZgt{}}

\PY{c}{\PYZsh{}\PYZlt{}listing\PYZhy{}2\PYZgt{}}
\PY{c}{\PYZsh{} ...}

\PY{k+kn}{from} \PY{n+nn}{numpy} \PY{k+kn}{import} \PY{n}{array}

\PY{n}{rhod} \PY{o}{=} \PY{n}{array}\PY{p}{(}\PY{p}{[}\PY{l+m+mf}{1.}   \PY{p}{]}\PY{p}{)}  
\PY{n}{th\PYZus{}d} \PY{o}{=} \PY{n}{array}\PY{p}{(}\PY{p}{[}\PY{l+m+mf}{305.} \PY{p}{]}\PY{p}{)}
\PY{n}{r\PYZus{}v}  \PY{o}{=} \PY{n}{array}\PY{p}{(}\PY{p}{[}\PY{l+m+mf}{0.01} \PY{p}{]}\PY{p}{)}  
\PY{n}{r\PYZus{}c}  \PY{o}{=} \PY{n}{array}\PY{p}{(}\PY{p}{[}\PY{l+m+mf}{0.001}\PY{p}{]}\PY{p}{)}
\PY{n}{r\PYZus{}r}  \PY{o}{=} \PY{n}{array}\PY{p}{(}\PY{p}{[}\PY{l+m+mf}{0.001}\PY{p}{]}\PY{p}{)}  
\PY{n}{dt}   \PY{o}{=} \PY{l+m+mi}{1} 

\PY{n}{blk\PYZus{}1m}\PY{o}{.}\PY{n}{adj\PYZus{}cellwise}\PY{p}{(}\PY{n}{opts}\PY{p}{,} 
  \PY{n}{rhod}\PY{p}{,}                \PY{c}{\PYZsh{} array, read\PYZhy{}only}
  \PY{n}{th\PYZus{}d}\PY{p}{,} \PY{n}{r\PYZus{}v}\PY{p}{,} \PY{n}{r\PYZus{}c}\PY{p}{,} \PY{n}{r\PYZus{}r}\PY{p}{,} \PY{c}{\PYZsh{} arrays, read\PYZhy{}write}
  \PY{n}{dt}                   \PY{c}{\PYZsh{} scalar}
\PY{p}{)}
\PY{k}{print} \PY{l+s}{\PYZdq{}}\PY{l+s}{r\PYZus{}v:}\PY{l+s}{\PYZdq{}}\PY{p}{,} \PY{n}{r\PYZus{}v}\PY{p}{,} \PY{l+s}{\PYZdq{}}\PY{l+s}{r\PYZus{}c:}\PY{l+s}{\PYZdq{}}\PY{p}{,} \PY{n}{r\PYZus{}c}\PY{p}{,} \PY{l+s}{\PYZdq{}}\PY{l+s}{r\PYZus{}r:}\PY{l+s}{\PYZdq{}}\PY{p}{,} \PY{n}{r\PYZus{}r}

\PY{c}{\PYZsh{} ...}
\PY{c}{\PYZsh{}\PYZlt{}/listing\PYZhy{}2\PYZgt{}}
\end{Verbatim}
  \vspace{-1.5em}%

    \label{lst:blk_1m_opts}
  \end{Listing}

  Among the options, there are Boolean flags for toggling condensation (\prog{cond}),
    cloud-water evaporation (\prog{cevp}), rain-water evaporation (\prog{revp}),
    autoconversion of cloud water into rain (\prog{conv}), accretion
    of cloud water by rain (\prog{accr}) and sedimentation of rain 
    (\prog{sedi}).
  There \prog{r\_c0} controls the threshold on cloud
    water mixing ratio above which autoconversion begins.
  The \prog{r\_eps} is~the~absolute tolerance in terms of mass mixing ratio
    that controls~the number of iterations within the saturation adjustment procedure.
  For further details, see description of Listing~3.1 in~\cite{Arabas_et_al_2015}.
  An instance of \prog{opts\_t} is expected as the first argument of~the 
    functions that constitute the interface of~the~single-moment scheme.

  The \prog{blk\_1m.adj\_cellwise()} function implements the
    saturation adjustment procedure.
  It models the condensational growth of cloud droplets as well as the
    evaporation of~water from cloud droplets and rain drops.
  An example calling sequence is presented in Listing~\ref{lst:blk_1m_adj}
    which is assumed to be a continuation of the code in Listing~\ref{lst:blk_1m_opts}.
  The \prog{adj\_cellwise()} function expects: an instance of \prog{opts\_t},
    five {\em NumPy} arrays of the model variables and a value of timestep \prog{dt}.
  The five arrays are expected to~contain values of~the~dry-air density $\rho_d$,
    that is not altered in the call, and four model state variables $\theta_d$, $r_v$, $r_c$, $r_r$.
  A call to~\prog{adj\_cellwise()} modifies the state variables.
  In the presented example, the arrays have only one element what corresponds
    to a parcel-model set-up.
  If multi-element arrays are given, the saturation adjustment procedure is applied 
    on each element.
  \begin{Listing}
  \refstepcounter{Listing}%
  \fvset{label=Listing~\theListing}%
  \fvset{gobble=0}%
  \renewcommand*\FancyVerbStartString{\PY{c}{\PYZsh{}\PYZlt{}listing\PYZhy{}2\PYZgt{}}}%
  \renewcommand*\FancyVerbStopString{\PY{c}{\PYZsh{}\PYZlt{}/listing\PYZhy{}2\PYZgt{}}}%
  \vspace{-1.5em}%

    \label{lst:blk_1m_adj}
  \end{Listing}

  The \prog{rhs\_cellwise()} and \prog{rhs\_columnwise()} functions implement
    representation of the collision-coalescence and the rain-sedimentation processes,
    respectively.
  They both takes five arguments, and they both expect the first one
    to be an instance of \prog{opts\_t}.

  For \prog{rhs\_cellwise()}, the next two arguments are {\em NumPy} arrays 
    to which the tendencies of cloud- and rain-water mixing ratio will be added.
  The last two arguments are {\em NumPy} arrays storing cloud- and
    rain-water mixing ratios; these will not be modified.

  For \prog{rhs\_columnwise()}, the second argument is a {\em NumPy} array 
    to which the tendency of rain-water mixing ratio will be~added.
  The third argument is expected to be a {\em NumPy} array with values of dry-air
    density.
  The array can have the same shape as other arrays, or can be a 
    single-column array.
  The fourth argument is expected to be a {\em NumPy} array of rain-water mixing ratios
    and will not be modified.
  The last argument is the vertical grid spacing.
  The second and the third dimensions are treated as the vertical ones
    for 2-D and 3-D arrays, respectively.
  It~is~assumed that the indices of the array increase with height.

%

  \section{Accessing {\em libcloudph++} from Fortran via Python}\label{sec:fortran}

  Python is an efficient ``glue'' language
    for coupling codes written in different programming languages.
  In this section, we present an~example solution for accessing {\em libcloudph++}
    from Fortran using the Python bindings.
  It is implemented using the \prog{CFFI}\footnote{\url{http://cffi.readthedocs.org/}} 
    (C Foreign Function Interface) Python package and 
    the \prog{ISO\_C\_BINDING} module that is part of the Fortran 2003 standard.
  Despite the fact that both \prog{CFFI} and \prog{ISO\_C\_BINDING} are intended for
    interfacing code written in the C language, the presented solution does not require a C compiler.
  
  \begin{Listing}
  \refstepcounter{Listing}%
  \fvset{label=Listing~\theListing (test.py)}%
  \fvset{gobble=0}%
  \renewcommand*\FancyVerbStartString{\PY{c}{\PYZsh{}\PYZlt{}listing\PYZhy{}1\PYZgt{}}}%
  \renewcommand*\FancyVerbStopString{\PY{c}{\PYZsh{}\PYZlt{}/listing\PYZhy{}1\PYZgt{}}}%
  \begin{Verbatim}[commandchars=\\\{\}]
\PY{c}{\PYZsh{}\PYZlt{}listing\PYZhy{}1\PYZgt{}}
\PY{k+kn}{from} \PY{n+nn}{cffi} \PY{k+kn}{import} \PY{n}{FFI}
\PY{k+kn}{from} \PY{n+nn}{libcloudphxx} \PY{k+kn}{import} \PY{n}{common}

\PY{n}{ffi} \PY{o}{=} \PY{n}{FFI}\PY{p}{(}\PY{p}{)}
\PY{n}{lib} \PY{o}{=} \PY{n}{ffi}\PY{o}{.}\PY{n}{dlopen}\PY{p}{(}\PY{l+s}{\PYZdq{}}\PY{l+s}{test.so}\PY{l+s}{\PYZdq{}}\PY{p}{)}
\PY{n}{ffi}\PY{o}{.}\PY{n}{cdef}\PY{p}{(}\PY{l+s}{\PYZdq{}}\PY{l+s}{void main(void*,void*);}\PY{l+s}{\PYZdq{}}\PY{p}{)}

\PY{n+nd}{@ffi.callback}\PY{p}{(}\PY{l+s}{\PYZdq{}}\PY{l+s}{double(double,double,double)}\PY{l+s}{\PYZdq{}}\PY{p}{)}
\PY{k}{def} \PY{n+nf}{rw3\PYZus{}cr}\PY{p}{(}\PY{n}{rd3}\PY{p}{,} \PY{n}{kappa}\PY{p}{,} \PY{n}{T}\PY{p}{)}\PY{p}{:}
  \PY{k}{return} \PY{n}{common}\PY{o}{.}\PY{n}{rw3\PYZus{}cr}\PY{p}{(}\PY{n}{rd3}\PY{p}{,} \PY{n}{kappa}\PY{p}{,} \PY{n}{T}\PY{p}{)}

\PY{n+nd}{@ffi.callback}\PY{p}{(}\PY{l+s}{\PYZdq{}}\PY{l+s}{double(double,double,double)}\PY{l+s}{\PYZdq{}}\PY{p}{)}
\PY{k}{def} \PY{n+nf}{S\PYZus{}cr}\PY{p}{(}\PY{n}{rd3}\PY{p}{,} \PY{n}{kappa}\PY{p}{,} \PY{n}{T}\PY{p}{)}\PY{p}{:}
  \PY{k}{return} \PY{n}{common}\PY{o}{.}\PY{n}{S\PYZus{}cr}\PY{p}{(}\PY{n}{rd3}\PY{p}{,} \PY{n}{kappa}\PY{p}{,} \PY{n}{T}\PY{p}{)}

\PY{n}{lib}\PY{o}{.}\PY{n}{main}\PY{p}{(}\PY{n}{rw3\PYZus{}cr}\PY{p}{,} \PY{n}{S\PYZus{}cr}\PY{p}{)}
\PY{c}{\PYZsh{}\PYZlt{}/listing\PYZhy{}1\PYZgt{}}
\end{Verbatim}
  \vspace{-1.5em}%

    \label{lst:test.py}
  \end{Listing}

  The code exemplifying the use of {\em libcloudph++} from Fortran
    is~presented in Listings \ref{lst:test.py} and \ref{lst:test.f}.
  The code prints the same table of critical radii and~supersaturations
    as the one from Listing~\ref{lst:kappa_crit} but doing the calls to 
    {\em libcloudph++} from Fortran.

  \begin{Listing}
  \refstepcounter{Listing}%
  \fvset{label=Listing~\theListing (test.f)}%
  \fvset{gobble=0}%
  \renewcommand*\FancyVerbStartString{\PY{c}{!\PYZlt{}listing\PYZhy{}1\PYZgt{}}}%
  \renewcommand*\FancyVerbStopString{\PY{c}{!\PYZlt{}/listing\PYZhy{}1\PYZgt{}}}%
  \begin{Verbatim}[commandchars=\\\{\}]
\PY{c}{!\PYZlt{}listing\PYZhy{}1\PYZgt{}}
\PY{k}{module }\PY{n+nv}{test}
  \PY{k}{interface}
\PY{k}{    }\PY{k}{function }\PY{n+nv}{f3arg}\PY{p}{(}\PY{n+nv}{a1}\PY{p}{,}\PY{n+nv}{a2}\PY{p}{,}\PY{n+nv}{a3}\PY{p}{)} \PY{k}{bind}\PY{p}{(}\PY{n+nv}{c}\PY{p}{)}
      \PY{k}{use }\PY{n+nb}{iso\PYZus{}c\PYZus{}binding}
\PY{n+nb}{      }\PY{k+kt}{real}\PY{p}{(}\PY{k+kt}{c\PYZus{}double}\PY{p}{)} \PY{k+kd}{::} \PY{n+nv}{f3arg}
      \PY{k+kt}{real}\PY{p}{(}\PY{k+kt}{c\PYZus{}double}\PY{p}{)}\PY{p}{,} \PY{k}{value} \PY{k+kd}{::} \PY{n+nv}{a1}\PY{p}{,}\PY{n+nv}{a2}\PY{p}{,}\PY{n+nv}{a3}
    \PY{k}{end}
\PY{k}{  }\PY{k}{end }\PY{k}{interface}

\PY{k}{  }\PY{k}{contains}

\PY{k}{  }\PY{k}{subroutine }\PY{n+nv}{main}\PY{p}{(}\PY{n+nv}{rw3\PYZus{}cr\PYZus{}p}\PY{p}{,} \PY{n+nv}{S\PYZus{}cr\PYZus{}p}\PY{p}{)} \PY{k}{bind}\PY{p}{(}\PY{n+nv}{c}\PY{p}{)}
    \PY{k}{use }\PY{n+nb}{iso\PYZus{}c\PYZus{}binding}
\PY{n+nb}{    }\PY{k}{type}\PY{p}{(}\PY{k+kt}{c\PYZus{}funptr}\PY{p}{)}\PY{p}{,} \PY{k}{value} \PY{k+kd}{::} \PY{n+nv}{rw3\PYZus{}cr\PYZus{}p}\PY{p}{,} \PY{n+nv}{S\PYZus{}cr\PYZus{}p}

    \PY{n+nv}{procedure}\PY{p}{(}\PY{n+nv}{f3arg}\PY{p}{)}\PY{p}{,} \PY{k}{pointer} \PY{k+kd}{::} \PY{n+nv}{rw3\PYZus{}cr}\PY{p}{,} \PY{n+nv}{S\PYZus{}cr}
    \PY{k}{call }\PY{n+nv}{c\PYZus{}f\PYZus{}procpointer}\PY{p}{(}\PY{n+nv}{rw3\PYZus{}cr\PYZus{}p}\PY{p}{,} \PY{n+nv}{rw3\PYZus{}cr}\PY{p}{)}
    \PY{k}{call }\PY{n+nv}{c\PYZus{}f\PYZus{}procpointer}\PY{p}{(}\PY{n+nv}{S\PYZus{}cr\PYZus{}p}\PY{p}{,} \PY{n+nv}{S\PYZus{}cr}\PY{p}{)}

    \PY{k}{block}
\PY{k}{      }\PY{k+kt}{real}\PY{p}{,} \PY{k}{dimension}\PY{p}{(}\PY{l+m+mi}{5}\PY{p}{)} \PY{k+kd}{::} \PY{n+nv}{rd} \PY{o}{=} \PY{p}{(}\PY{o}{/}\PY{l+m+mf}{.0223}\PY{p}{,} \PY{l+m+mf}{.0479}\PY{p}{,} \PY{l+m+mf}{.103}\PY{p}{,} \PY{l+m+mf}{.223}\PY{p}{,} \PY{l+m+mf}{.479}\PY{o}{/}\PY{p}{)}
      \PY{k+kt}{real}\PY{p}{(}\PY{k+kt}{c\PYZus{}double}\PY{p}{)} \PY{k+kd}{::} \PY{n+nv}{kappa} \PY{o}{=} \PY{l+m+mf}{1.28}\PY{p}{,} \PY{n+nv}{T} \PY{o}{=} \PY{l+m+mi}{273}\PY{p}{,} \PY{n+nv}{rd3}
      \PY{k+kt}{integer} \PY{k+kd}{::} \PY{n+nv}{i}

      \PY{k}{print} \PY{l+s+s1}{\PYZsq{}(3A10)\PYZsq{}}\PY{p}{,} \PY{l+s+s1}{\PYZsq{}rd [um]\PYZsq{}}\PY{p}{,} \PY{l+s+s1}{\PYZsq{}r* [um]\PYZsq{}}\PY{p}{,} \PY{l+s+s1}{\PYZsq{}S*\PYZhy{}1 [\PYZpc{}]\PYZsq{}}
      \PY{k}{do }\PY{n+nv}{i}\PY{o}{=}\PY{l+m+mi}{1}\PY{p}{,} \PY{n+nv}{size}\PY{p}{(}\PY{n+nv}{rd}\PY{p}{)}
        \PY{n+nv}{rd3} \PY{o}{=} \PY{p}{(}\PY{n+nv}{rd}\PY{p}{(}\PY{n+nv}{i}\PY{p}{)} \PY{o}{*} \PY{l+m+mi}{1e\PYZhy{}6}\PY{p}{)}\PY{o}{**}\PY{l+m+mi}{3}
        \PY{k}{print} \PY{l+s+s1}{\PYZsq{}(3G10.2)\PYZsq{}}\PY{p}{,}                      \PY{p}{\PYZam{}}
          \PY{n+nv}{rd}\PY{p}{(}\PY{n+nv}{i}\PY{p}{)}\PY{p}{,}                               \PY{p}{\PYZam{}}
          \PY{p}{(}\PY{n+nv}{rw3\PYZus{}cr}\PY{p}{(}\PY{n+nv}{rd3}\PY{p}{,} \PY{n+nv}{kappa}\PY{p}{,} \PY{n+nv}{T}\PY{p}{)}\PY{o}{**}\PY{p}{(}\PY{l+m+mi}{1}\PY{o}{/}\PY{l+m+mf}{3.}\PY{p}{)}\PY{p}{)}\PY{o}{*}\PY{1+m+mf}{1e6}\PY{p}{,}  \PY{p}{\PYZam{}}
          \PY{p}{(}\PY{n+nv}{S\PYZus{}cr}\PY{p}{(}\PY{n+nv}{rd3}\PY{p}{,} \PY{n+nv}{kappa}\PY{p}{,} \PY{n+nv}{T}\PY{p}{)} \PY{o}{\PYZhy{}} \PY{l+m+mi}{1}\PY{p}{)} \PY{o}{*} \PY{l+m+mi}{100}
      \PY{n+nv}{enddo}
    \PY{k}{end }\PY{k}{block}
\PY{k}{  }\PY{k}{end}
\PY{k}{end}
\PY{c}{!\PYZlt{}/listing\PYZhy{}1\PYZgt{}}
\end{Verbatim}
  \vspace{-1.5em}%

    \label{lst:test.f}
  \end{Listing}

  The central idea is to provide a common addressing space for Python and Fortran,
    so that it is possible to refer to the Python code from Fortran and vice versa.
  To implement such C++/Python/Fortran coupling, the Fortran code is compiled 
    as~a~shared library to be loaded from Python.

  Example commands to compile and execute the codes from Listings~\ref{lst:test.py}
    and~\ref{lst:test.f} are presented in Listing~\ref{lst:make.sh}.
  \begin{Listing}
  \refstepcounter{Listing}%
  \fvset{label=Listing~\theListing}%
  \fvset{gobble=0}%
  \renewcommand*\FancyVerbStartString{\PY{c}{\PYZsh{}\PYZlt{}listing\PYZhy{}1\PYZgt{}}}%
  \renewcommand*\FancyVerbStopString{\PY{c}{\PYZsh{}\PYZlt{}/listing\PYZhy{}1\PYZgt{}}}%
  \begin{Verbatim}[commandchars=\\\{\}]
\PY{c}{\PYZsh{}\PYZlt{}listing\PYZhy{}1\PYZgt{}}
gfortran \PYZhy{}ffree\PYZhy{}form \PYZhy{}std\PY{o}{=}f2008 \PYZhy{}shared \PYZhy{}fPIC test.f \PYZhy{}o test.so
\PY{n+nv}{LD\PYZus{}LIBRARY\PYZus{}PATH}\PY{o}{=}. python test.py
\PY{c}{\PYZsh{}\PYZlt{}/listing\PYZhy{}1\PYZgt{}}
\end{Verbatim}
  \vspace{-1.5em}%

    \label{lst:make.sh}
  \end{Listing}
  First, the \prog{gfortran} compiler is instructed to compile \prog{test.f}
    into a shared library \prog{test.so}.
  Second, the Python script is run being instructed to~include the current directory
    (``\prog{.}'') in the shared-library search path. 
  Consequently, even though the goal is to call Python from Fortran, the control flow 
    starts in Python. 

  In the Python code in Listing~\ref{lst:test.py},
    the \prog{test.so} library is~loaded by using the \prog{CFFI}'s 
    interface to the \prog{dlopen()} system call.
  In~order to enable accessing the Fortran subroutine \prog{main}, 
    the datatypes of the arguments of \prog{main} are specified using 
    the \prog{CFFI}'s \prog{cdef} function.
  The \prog{void*} pointers are interpreted
    as function pointers within \prog{main}.
  The \prog{CFFI} ``callback'' mechanism is used to~create callback
    objects for which Fortran-compatible function pointers can be obtained.
  The callback objects are defined as functions with the definition prepended with 
    a \prog{@ffi.callback} decorator.
  The~decorator includes the C signature of the function.
  The~signature specifies the datatypes of the return value and~of~the~arguments.
  The callback objects defined in this way are then passed as arguments
    to \prog{main}.

  In the Fortran code in Listing~\ref{lst:test.f}, the definition
    of subroutine \prog{main} is preceded by a definition of an interface
    to a three-argument function.
  Both the interface and the \prog{main} subroutine are assigned with the
    \prog{bind(c)} attribute which ensures \prog{CFFI}-compatible naming
    of functions in the shared library.
  The arguments to \prog{main} are defined as C function pointers. 
  Within \prog{main}, the three-argument function interface is used
    to define two Fortran function pointers \prog{rw3\_cr} and \prog{S\_cr}.
  The Fortran function pointers are associated with the C ones 
    by calling the \prog{c\_f\_procpointer} subroutine 
    (defined in the standard \prog{ISO\_C\_BINDING} module).
  Afterwards, the~\prog{rw3\_cr} and \prog{S\_cr} can be used as any other function in Fortran.

  Output from execution of the two commands from Listing~\ref{lst:make.sh}
    is presented in Listing~\ref{lst:f_out}.
  It matches the result obtained with Python code from listing~\ref{lst:kappa_crit}
    presented in Listing~\ref{lst:kappa_crit_out}.

  \begin{Listing}
  \refstepcounter{Listing}%
  \fvset{label=Listing~\theListing}%
  \fvset{gobble=0}%
  \renewcommand*\FancyVerbStartString{\PYZsh{}\PYZlt{}listing\PYZhy{}1\PYZgt{}}%
  \renewcommand*\FancyVerbStopString{\PYZsh{}\PYZlt{}/listing\PYZhy{}1\PYZgt{}}%
  \begin{Verbatim}[commandchars=\\\{\}]
\PYZsh{}\PYZlt{}listing\PYZhy{}1\PYZgt{}
   rd [um]   r* [um]  S*\PYZhy{}1 [\PYZpc{}]
  0.22E\PYZhy{}01  0.19      0.39    
  0.48E\PYZhy{}01  0.61      0.13    
  0.10       1.9      0.40E\PYZhy{}01
  0.22       6.1      0.12E\PYZhy{}01
  0.48       19.      0.40E\PYZhy{}02
\PYZsh{}\PYZlt{}/listing\PYZhy{}1\PYZgt{}
\end{Verbatim}
  \vspace{-1.5em}%

    \label{lst:f_out}
  \end{Listing}

  \section{Remarks}
  
  The presented bindings to {\em libcloudph++} provide access 
    to~the~library from Python, a de-facto standard interpreted language in~science.
  Availability of the bindings enlarges the potential user~base of~{\em libcloudph++}.
  It also enlarges the range of the library applications by offering
    the possibility to couple {\em libcloudph++} with numerous existing
    Python packages, e.g. with \prog{SciPy} or~\prog{CFFI} as exemplified
    in this paper.

  The presented C++/Python/Fortran coupling method is,~to~the authors' knowledge, a novel approach.
  While not being straightforward, it has the advantage of offering 
    productivity-oriented features of an interpreted language,
    even though the addressed problem concerns coupling of two compiled
    languages.
  
  \appendix

  \section{Obtaining and installing the bindings}\label{sec:install}

  The Python bindings for {\em libcloudph++} are shipped with the~library.
  The library uses {\em CMake}\footnote{\url{http://cmake.org/}} for build and test automation.
  Besides {\em CMake}, the library code depends on several components 
    of~the~{\em Boost} C++ library collection and on the Thrust\footnote{\url{https://thrust.github.io/}} C++ library.
  If~available, the library will be compiled
    with support for~parallelisation of the particle-based algorithm
    using OpenMP and~CUDA.

  Listing~\ref{lst:install.sh} gives a set of commands that result in downloading the~current development
    version of the library, compilation, execution of test programs, and
    installation of the library and Python bindings on the system.
  For reference on how to use non-default compiler or how to use non-default paths, see 
    documentation of~{\em CMake} and the README file shipped with {\em libcloudph++}.
  \begin{Listing}
  \refstepcounter{Listing}%
  \fvset{label=Listing~\theListing}%
  \fvset{gobble=0}%
  \renewcommand*\FancyVerbStartString{\PYZsh{}\PYZlt{}listing\PYZhy{}1\PYZgt{}}%
  \renewcommand*\FancyVerbStopString{\PYZsh{}\PYZlt{}/listing\PYZhy{}1\PYZgt{}}%
  \begin{Verbatim}[commandchars=\\\{\}]
\PYZsh{}\PYZlt{}listing\PYZhy{}1\PYZgt{}
git clone http://github.com/igfuw/libcloudphxx
cd libcloudphxx
mkdir build; cd build
cmake ..
make all test
sudo make install
\PYZsh{}\PYZlt{}/listing\PYZhy{}1\PYZgt{}
\end{Verbatim}
  \vspace{-1.5em}%

    \label{lst:install.sh}
  \end{Listing}
  
  \section*{Acknowledgements}
  \footnotesize

  We thank Maciej Fijałkowski for his help with Python and CFFI.
  We acknowledge comments to the initial version of the manuscript from
    Anna Jaruga.
  DJ acknowledges support from the Polish Ministry of Science and Higher Education 
    (project no.~1119/MOB/13/2014/0). 
  SA acknowledges support from Poland’s National Science Centre (Narodowe Centrum Nauki) 
    [decision no.~2012/06/M/ST10/00434 (HARMONIA)].
  The National Center for Atmospheric Research (NCAR) is sponsored by the National Science Foundation. 

  \renewcommand*{\bibfont}{\footnotesize}
  \printbibliography

\end{document}